\documentclass[twocolumn,showpacs,preprintnumbers,amsmath,amssymb]{revtex4-1}
\newcommand{\be}{\begin{equation}}
\newcommand{\ee}{\end{equation}}
\newcommand{\mc}{\mathcal}
\usepackage{lineno}
\usepackage[dvips]{graphicx}
\usepackage{color} 
\usepackage{epsfig}
\usepackage{soul}
\begin{document}

\title{Calculation of quantum discord for qubit-qudit or $N$ qubits}

\author{Sai Vinjanampathy$^{1}$, and A. R. P. Rau$^2$}
\affiliation{
$^1$Department of Physics, University of Massachusetts at Boston, Boston, MA 02125, USA\\
$^2$Department of Physics and Astronomy, Louisiana State University, Baton Rouge, Louisiana 70803, USA}
\begin{abstract}
Quantum discord, a kind of quantum correlation, is defined as the difference between quantum mutual information and classical correlation in a bipartite system. It has been discussed so far for small systems with only a few independent parameters. We extend here to a much broader class of states when the second party is of arbitrary dimension $d$, so long as the first, measured, party is a qubit. We present two formul{\ae} to calculate quantum discord, the first relating to the original entropic definition and the second to a recently proposed geometric distance measure which leads to an analytical formulation. The tracing over the qubit in the entropic calculation is reduced to a very simple prescription. And, when the $d$-dimensional system is a so-called $X$ state, the density matrix having non-zero elements only along the diagonal and anti-diagonal so as to appear visually like the letter X, the entropic calculation can be carried out analytically. Such states of the full bipartite qubit-qudit system may be named ``extended $X$ states", whose density matrix is built of four block matrices, each visually appearing as an X. The optimization involved in the entropic calculation is generally over two parameters, reducing to one for many cases, and avoided altogether for an overwhelmingly large set of density matrices as our numerical investigations demonstrate.  

Our results also apply to states of a $N$-qubit system, where ``extended $X$ states" consist of $(2^{N+2}-1)$ states, larger in number than the $(2^{N+1}-1)$ of $X$ states of $N$ qubits. While these are still smaller than the total number $(2^{2N}-1)$ of states of $N$ qubits, the number of parameters involved is nevertheless large. In the case of $N=2$, they encompass the entire 15-dimensional parameter space, that is, the extended $X$ states for $N=2$ represent the full qubit-qubit system.  

\end{abstract}


\maketitle
\section{Introduction}
Quantum correlations in a bipartite system AB have generated interest for various tasks such as computing \cite{ref1}, imaging \cite{ref1p1} and metrology \cite{ref1p2}. Like entanglement, which has been proven to be a resource for such tasks \cite{ref1}, quantum discord \cite{ref2} was proposed to be a resource behind the efficiency of the so-called DQC-1 model \cite{ref3}. Though this has since been disputed \cite{ref19}, it is of interest to understand quantum discord and its potential usefulness in quantum technologies.

Reliable and convenient measures of entanglement, such as concurrence \cite{ref4} and positive partial transpose \cite{ref5}, exist only for AB a qubit-qubit or qubit-qutrit system \cite{ref6}. Quantum discord, on the other hand, is calculated very differently, through computation of mutual information as captured by Shannon entropy \cite{ref2}. While the calculation of eigenvalues and thereby entropy is relatively straightforward, a main difficulty in computing quantum discord is to calculate the classical correlation, discord being defined by default as all that remains of the quantum mutual information between A and B after subtracting out the classical correlation between them. This calculation involves considering all possible measurements on one of the subsystems, say A, tracing over A, and then computing the entropy which is subtracted from the quantum mutual information of AB. This requirement of considering all possible measurements so as to account for any classical correlation that may exist between A and B calls for an extremization procedure that can be tedious, especially with increasing parameters. Only a few results are available with, in particular, hardly any analytical procedures worked out for general systems. For qubit-qubit, the entire space being of 15 parameters, results \cite{ref7} are available for a class of 3-parameter states that were recently extended by one of us and coworkers to 7 \cite{ref8}. This class of states have been called $X$ states \cite{ref9} because they have non-zero entries only along the diagonal and anti-diagonal of the $4 \times 4$ density matrix.
 
$X$ states of two qubits have already found application in many studies of entanglement \cite{ref9} and discord \cite{ref10}. They have the virtue that many calculations can be carried out analytically which is always an aide to understanding and application. At the same time, they encompass a variety of states of interest, separable and nonseparable, classical and non-classical, so that confining our studies to them is not unduly restrictive. Recently, the algebraic symmetry structure of $X$ states has been revealed \cite{ref11} and we have extended the description to $X$ states of $N$ qubits \cite{ref12}. Again, these $X$ states embrace many different states of interest such as W \cite{ref13} and GHZ  \cite{ref14} states of three qubits and Dicke states of general $N$ \cite{ref15}.

In this paper, we show how our previous procedure for calculating discord for $N=2$ $X$ states, both the tracing over the first subsystem A and the extremization, extends rather readily to general $N$; the final step of extremization is at most over two parameters and even that can be avoided in many cases by straightforward algebraic evaluation at a few values of those parameters. The key observation is that even though there is an explosion in the number of parameters contained in the density matrix at larger $N$, there being $(2^{N+1} -1)$ of them, the $X$ structure breaks into calculations of a series of $2 \times 2$ matrices whose eigenvalues can be written down in closed form easily. Further, so long as A is a qubit, no matter what the dimensionality of the second subsystem B, the extremization over the parameters involved in von Neumann measurements on it, is also limited and standard and can be rendered in a ``universal"  closed form, the same that was already encountered in $N=2$ \cite{ref8}. Indeed, our results hold even beyond $X$ states, applying to any AB of the form qubit-qudit, however large the dimension $d$ of B, so long as the density matrix of AB, when written as four $d \times d$ block matrices, has $X$ character for each of these blocks. The density matrix $\rho_{AB}$ need not itself have $X$ character. Such states may be termed ``extended $X$", $\rho_{AB}$ having the form $2 \otimes d_X$, the $d$-dimensional part being an $X$ state. The calculation of eigenvalues for the reduced density matrix $\rho_B$ after tracing over A breaks into that of a series of $2 \times 2$ matrices and of the full density matrix $\rho_{AB}$ to at most $4 \times 4$ matrices, all of which can be obtained analytically.

The outline of the paper is as follows. In Section \ref{s2} we present the definition of generalized $X$-states for $N$-qubits and of extended $X$-states. In Section \ref{s3}, we will outline two formul{\ae}, an entropy-based formula and a distance formula for calculating quantum discord. We will discuss the reduction of the optimization from a two-parameter optimization in the general case to a single parameter optimization under certain circumstances and also point out when an explicit analytical form can be written down for the entropic formula for quantum discord. These parameters either define the measurement basis used to compute the so-called classical correlations in the entropic formula or define the classical states $\mathrm{C}$ of zero quantum discord for the geometric distance measure. Finally, we present our conclusions in Section \ref{s4}.

\section{$X$ states of qubit-qudit and $N$-qubit systems}\label{s2}
$X$ states were originally described \cite{ref9} for qubit-qubit systems as those with non-zero density matrix elements only along the diagonal and anti-diagonal. The former's three real and the latter's two complex parameters form a total of seven independent parameters to describe such a state. Later, an underlying symmetry structure was pointed out, that these states conform to a su(2) $\times$ su(2) $\times$ u(1) sub-algebra of the full su(4) algebra of a qubit-qubit system \cite{ref11}. Thus, other 7-parameter density matrices that do not have the visual structure of the letter X but share the same algebra belong to the same category. This recognition also led to a natural generalization to $N$-qubit systems \cite{ref12}. Starting from a single qubit, when all states are trivially $X$ and have the su(2) algebra, each successive $N$ is generated by doubling and adding a u(1). For $N=2$, we have su(2) $\times$ u(1) $\times$ su(2), and upon going to $N=3$, the symmetry is of su(2) $\times$ u(1) $\times$ su(2) $\times$ u(1) $\times$ su(2) $\times$ u(1) $\times$ su(2), a sub-algebra of su(8). The sub-algebra deals with a 15-parameter set of states. There are $(2^{N+1} -1)$ parameters for general $N$. 

Similarly, for a qubit-qudit system, there are $(4d-1)$ $X$ states. Given an underlying sub-algebra of states and operators, so long as all operators on the system including the Hamiltonian and even possibly those leading to dissipation and decoherence lie within that sub-algebra, the evolution stays within $X$ character. And, these actions can be calculated analytically. For this reason, $X$ states have figured in a wide variety of studies in quantum information. While not the full system of $N$ qubits or of qubit-qudit, they are nevertheless a large parameter set that is capable of describing a variety of systems and phenomena of interest while remaining tractable.

For purposes of the next section, where we are interested in tracing over one qubit, we generalize to an even larger category of states. Reduced density matrices obtained from such bipartite states have the $X$ character, while the bipartite states themselves do not.  We will refer to such states as ``extended $X$ states". In a bipartite qubit-qudit system, for $d$ odd, these involve $(8d-5)$ and, for $d$ even, $(8d-1)$ parameters. For $N$ qubits, with $d=2^{N-1}$, there are $(2^{N+2} -1)$  states. These numbers are larger than the corresponding ones for $X$ states. While still smaller than $(4d^2-1)$ or $(2^{2N}-1)$ that represent the total systems in general, they are nevertheless substantial and, for $N=2$, actually coincide with the total of 15 parameters, thus applying to all possible qubit-qubit states. For this important system, therefore, our analytical result applies to a general density matrix with no restriction on its form or its individual elements.

The non-zero elements of $N$-qubit $X$ states are $\rho_{ii}$ and $\rho_{i,2^N+1-i}$ for $i=1, \ldots ,N$. For extended $X$ states, in addition, $\rho_{i, 2^{N-1}+i}$ and $\rho_{i,2^{N-1}+1-i}$ for $i=1, \ldots ,N/2$ and $\rho_{i,i-2^{N-1}}$ and $\rho_{i,2^N+1-(i-2^{N-1})}$ for $i=N/2+1, \ldots ,N$ are non-zero. Viewed as four equal blocks, the density matrix has each block appearing as an X. An example for $N=3$ is 

\begin{equation}
\rho =  \left( 
\begin{array}{cccccccc}
\rho_{11} & 0 & 0 & \rho_{14} & \rho_{15} & 0 & 0 & \rho_{18} \\ 
0 & \rho_{22} & \rho_{23} & 0 & 0 & \rho_{26} & \rho_{27} & 0 \\ 
0 & \rho_{32} & \rho_{33} & 0 & 0 & \rho_{36} & \rho_{37} & 0 \\
\rho_{41} & 0 & 0 & \rho_{44} & \rho_{45} & 0 & 0 & \rho_{48} \\
\rho_{51} & 0 & 0 & \rho_{54} & \rho_{55} & 0 & 0 & \rho_{58} \\
0 & \rho_{62} & \rho_{63} & 0 & 0 & \rho_{66} & \rho_{67} & 0 \\
0 & \rho_{72} & \rho_{73} & 0 & 0 & \rho_{76} & \rho_{77} & 0 \\
\rho_{81} & 0 & 0 & \rho_{84} & \rho_{85} & 0 & 0 & \rho_{88} 
\end{array}
\right).      
\label{eqn1}
\end{equation}

In the next section, we will use these states to write down formul{\ae} for quantum discord explicitly.

\section{Two formul{\ae} for calculating quantum discord}\label{s3}
To establish the first formula, we note that quantum discord was originally defined entropically as \cite{ref2} 
\be\label{eqn2}
\mc{Q}(\rho^{AB})  := \mc{I}(\rho^{AB})-\mc{C}(\rho^{AB}),
\ee
where the quantum mutual information  is defined as $\mc{I}(\rho^{AB}):=S(\rho^{A})+S(\rho^{A})-S(\rho^{AB})$. Here $S(\rho)$ is the von-Neumann entropy. The second term in the definition is the classical correlation defined as 
\be\label{eqn3}
\mc{C}(\rho^{AB})=\displaystyle \sup_{A_i}\left(S(\rho^{B})-S(\rho^{AB}\vert\{A_{i}\})\right),
\ee 
where $S(\rho^{AB}\vert\{A_{i}\})=\displaystyle\sum_{i}p_{i}S(\rho_{i})$. Here
\be\label{eqn4}
\rho_{i}=\frac{A_{i}\rho^{AB}A_{i}}{p_i},
\ee 
where the measurement operators $A_{i}=U\vert i\rangle\langle i\vert U^{\dagger}$ are chosen by transforming the orthogonal projectors $\Pi_i = |i\rangle\langle i |, \, i=0,1$ for subsystem A along the computational basis kets $\vert i\rangle$ by a general unitary transformation $U$, and $p_{i}=\mathrm{tr}(A_{i}\rho^{AB}A_{i})$ \cite{ref7, ref8}. 

For single qubit measurements, the unitary operator $U$ can be parametrized in terms of the Pauli matrices as $U=t I+i\vec{y}.\vec{\sigma}$, where the four parameters $t,\vec{y}$ are constrained by unitarity: $t^2+y_1^2+y_2^2+y_3^2=1$ \cite{ref7}. The three independent parameters can be redefined into the unit vector $\vec{z}$ defined as \cite{ref7} 
\be\label{eqn5}
\vec{z}=\{2(-ty_2+y_1y_3),2(ty_1+y_2y_3),t^2+y_3^2-y_1^2-y_2^2\}.
\ee

This unit vector is relevant for the choice of measurement directions $i=\pm\hat{z}$ and alternative sets of the parameters $t,\vec{y}$ may be defined for other orthogonal directions \cite{ref7}. For any vector $\vec{V}$ and direction $\pm\hat{z}$, it is easy to establish the vector identity 
\be\label{eqn6}
A_{i}(\vec{\sigma}.\vec{V})A_i=\pm(\vec{z}.\vec{V})A_i.
\ee
Furthermore, $A_{i}=U\vert i\rangle\langle i\vert U^{\dagger}=(1\pm\vec{z}.\vec{\sigma})/2$. Hence, this can be rendered in the standard polar decomposition as
\be\label{eqn7}
A_{+}=
\left(
\begin{array}{ccc}
\cos^2(\frac{\theta}{2})  &  \frac{1}{2}\sin{\theta}e^{-i\phi} \\
 \frac{1}{2}\sin{\theta}e^{i\phi}  &     \sin^2(\frac{\theta}{2}) 
\end{array}
\right),
\ee
and $A_{-}$, its parity conjugate, is obtained by the substitution $(\theta,\phi)\rightarrow(\pi-\theta,\pi+\phi)$. 

All the algebra involved in calculating the post-measurement state in Eq. (\ref{eqn4}) can finally be reduced to the following simple prescription. Viewing a density matrix $\rho^{AB}$ such as in Eq. (\ref{eqn1}) as a block $2\times2$ matrix of four equal blocks, $A_{+}\rho^{AB}A_{+}$ is calculated by multiplying the four blocks, considered in a clockwise manner from the top left, by the elements in Eq. (\ref{eqn7}), that is, $\cos^{2}(\frac{\theta}{2})$, $\frac{1}{2}\sin{\theta}e^{-i\phi}$, $\sin^2(\frac{\theta}{2})$ and $\frac{1}{2}\sin{\theta}e^{i\phi}$, respectively. Tracing out subsystem $A$ means adding together the four blocks of this post-measurement density matrix. Note that since the eigenvectors of this matrix depend only on the unit vector $\vec{z}$, they are only functions of two parameters, the polar angles $(\theta,\phi)$ of $\vec{z}$. In the two-qubit case, these two parameters were alternatively understood as arising from four parameters $k,l,m$ and $n$ constrained by two equations $k+l=1$ and $m^2+n^2=klm$ \cite{ref8}.
 
Before examining the supremum in Eq. (\ref{eqn3}), we comment on the choice of extended $X$-states with respect to evaluating quantum discord. Note that there exist two separate issues that make the evaluation of quantum discord a difficult task. The first of these is the fact that since polynomials of degree $n>4$ are not generally solvable by radicals \cite{ref16}, analytical computation of eigenvalues for $2 \otimes d$ systems is not guaranteed for $d>2$ (for $N$ qubits, $d= 2^{N-1}$). However, this is possible when restricted to extended $X$-states, the eigenvalue equation involving at most a quartic polynomial. 

For $N=3$, the post-measurement state corresponding to $A_{+}$ up to normalization, namely $A_{+}\rho A_{+}$, is given by
 \begin{widetext}
 \be
\left(
\begin{array}{cccccccc}
 \rho_{11} \cos^2\left(\frac{\theta }{2}\right) & 0 & 0 & \rho_{14} \cos^2\left(\frac{\theta }{2}\right) & s  \rho_{15} \sin (\theta ) & 0 & 0 & s  \rho_{18} \sin (\theta ) \\
 0 & \rho_{22} \cos^2\left(\frac{\theta }{2}\right) & \rho_{23} \cos^2\left(\frac{\theta }{2}\right) & 0 & 0 & s  \rho_{26} \sin (\theta ) & s  \rho_{27} \sin (\theta ) & 0 \\
 0 & \rho_{32} \cos^2\left(\frac{\theta }{2}\right) & \rho_{33} \cos^2\left(\frac{\theta }{2}\right) & 0 & 0 & s  \rho_{36} \sin (\theta ) & s  \rho_{37} \sin (\theta ) & 0 \\
 \rho_{41} \cos^2\left(\frac{\theta }{2}\right) & 0 & 0 & \rho_{44} \cos^2\left(\frac{\theta }{2}\right) & s  \rho_{45} \sin (\theta ) & 0 & 0 & s  \rho_{48} \sin (\theta ) \\ 
 s^{*}  \rho_{51} \sin (\theta ) & 0 & 0 & s^{*}  \rho_{54} \sin (\theta ) & \rho_{55}
   \sin^2\left(\frac{\theta }{2}\right) & 0 & 0 & \rho_{58} \sin^2\left(\frac{\theta }{2}\right) \\
 0 & s^{*} \rho_{62} \sin (\theta ) & s^{*}  \rho_{63} \sin (\theta ) & 0 & 0 & \rho_{66}
   \sin^2\left(\frac{\theta }{2}\right) & \rho_{67} \sin^2\left(\frac{\theta }{2}\right) & 0 \\
 0 & s^{*}  \rho_{72} \sin (\theta ) & s^{*}  \rho_{73} \sin (\theta ) & 0 & 0 & \rho_{76}
   \sin^2\left(\frac{\theta }{2}\right) & \rho_{77} \sin^2\left(\frac{\theta }{2}\right) & 0 \\
 s^{*}  \rho_{81} \sin (\theta ) & 0 & 0 & s^{*}  \rho_{84} \sin (\theta ) & \rho_{85}
   \sin^2\left(\frac{\theta }{2}\right) & 0 & 0 & \rho_{88} \sin^2\left(\frac{\theta }{2}\right)
\end{array}
\right),
\ee
 \end{widetext}
where $s=e^{-i \phi}/2$. The post-measurement reduced density matrix for B, given by adding the four $4\times4$ blocks, is of course an $X$ state of two qubits whose eigenvalues are easily obtained. 

\begin{figure}[h!]  
\begin{center}  
\includegraphics[height=70mm,width=70mm]{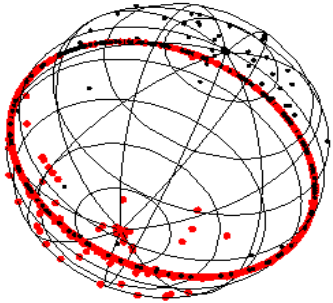}  
\caption{\small \sl Points correspond to the optimal measurement for 10,000 (Hilbert-Schmidt) randomly chosen two-qubit and three-qubit X-states obtained by full numerical optimization. To accommodate both plots on one Bloch sphere, we exploited parity symmetry which requires only half of the sphere, displaying $\{\theta,\phi\}$ for two qubit density matrices (thin, black) and $\{\theta+\pi/2,\phi+\pi\}$ for three-qubit density matrices (thick, red). For $\approx 99.8\%$  of randomly chosen two-qubit $X$-states and $\approx 99.5\%$ of randomly chosen three-qubit $X$-states, $\theta=0$ or $\theta=\pi/2$ provides the true discord. 
\label{Bloch}}  
\end{center}  
\end{figure}

  The second issue relates to the optimization involved in computing the classical correlations $\mathcal{C}(\rho)$ in Eq. (\ref{eqn3}). In evaluating the supremum, we note that the conditional entropy is symmetric in $i=\pm\hat{z}$. It is therefore symmetric under parity inversion of $\hat{z}$, namely $(\theta,\phi)\rightarrow(\pi-\theta,\pi+\phi)$. Therefore, the search for an optimum can be restricted to half the range of one of the angles, either to $0\leq\theta<\pi/2$ or to $0\leq\phi<\pi$. The proposal in \cite{ref8} of restricting to these extremal values which provides a formula for quantum discord for $N$-qubit $X$-states that is fully analytical and bypasses optimization works when there is only one extremum at $\theta=\pi/2$. But when there is more than one value for the extremum, they occur in pairs at possibly other values of the angles and an actual optimization is required. Interestingly, as noted in \cite{ref17}, such an optimization \textit{almost} always results in the solutions $\theta=0$ or $\theta=\pi/2$ and $\phi=0$ or $\phi=\pi$. We support this conclusion through our numerical investigation of quantum discord for three qubit $X$-states. We sampled 10,000 random density matrices chosen according to the Hilbert-Schmidt measure and constructed $X$-states using $\rho_X=\sum_{i}E_{i}\rho E^{\dagger}_{i}$, where $E_i$ are diagonal matrices given by $E_{1}=\mathrm{diag}(1 ,0, 0, 0, 0 ,0, 0 ,1)$, $E_{2}=\mathrm{diag}(0 ,1, 0, 0, 0, 0, 1, 0)$, $E_{3}=\mathrm{diag}(0, 0, 1, 0, 0, 1, 0, 0)$ and $E_{4}=\mathrm{diag}(0, 0, 0, 1, 1, 0, 0, 0)$. Discord was then computed by full numerical optimization and the optimal angles corresponding to each randomly generated state was plotted on the Bloch sphere. We note that for 99.5$\%$ of the states chosen this way, the optimal measurement corresponds to $\theta$ being at either poles or at the equator, as per \cite{ref8}. These optimal measurements are presented in figure (\ref{Bloch}).
Furthermore, when either the off-diagonals of the post-measurement density matrix are all the same, or when only one of the off-diagonals of the post-measurement density matrix is non-zero, it is easy to see that $\phi$ does not appear in the eigenvalues of the post-measurement density matrix and may be set arbitrarily at $\phi=0$. Hence, the optimization reduces to that in a single variable $\theta$, which is almost always either $\theta=0$ or $\theta=\pi/2$. See figure (\ref{Bloch}).

Finally, we turn to an alternative formula for quantum discord for $2 \otimes d$ systems, a geometric measure defined as \cite{ref18,ref19} 
\be\label{eqn9}
D^{(2)}_{A}(\rho)=\displaystyle\min_{\chi\in\mathcal{C}}\Vert\rho-\chi\Vert^2,
\ee
where $\chi$ is a classical state. Such a classical state, in general, can be written as
\be
\chi=\displaystyle\sum_{i=1}^{d_A}p_i\Pi^{(A)}_i\otimes\rho^{(B)}_{i},
\ee 
where $d_A$ is the dimensionality of subsystem A and $\Pi^{(A)}_{i}$ are its projectors. $\rho^{(B)}_i$ are density matrices describing states of subsystem B. For two-qubit systems, classical states are defined in terms of one $p_i$, two angles defining the projectors $\Pi^{(A)}_{i}$ and two sets of three parameters defining each $\rho_i$, a total of 9 parameters. 

The 15 parameters that describe any two-qubit state on the other hand, can be described in terms of three terms, two local Bloch vectors which are three dimensional and one $3\times3$ tensor. They are defined via the equations $x_{i}\equiv \mathrm{tr}(\rho*\sigma_{i}\otimes I)$, $y_{i}\equiv\mathrm{tr}(\rho*I\otimes \sigma_{i})$ and $T_{ij}\equiv\mathrm{tr}(\rho*\sigma_{i}\otimes \sigma_{j})$. These three tensors are defined for classical states via the equations $\vec{e}\equiv\mathrm{tr}(\langle\psi\vert\vec{\sigma}\vert\psi\rangle)$, $\vec{s}_{+}\equiv\mathrm{tr}((p_1\rho_1+p_2\rho_2)*\vec{\sigma})$ and $T_{ij}\equiv\mathrm{tr}(\rho*\sigma_{i}\otimes \sigma_{j})$. Since $\Vert\vec{e}\Vert=1$, there is another independent parameter, namely $t\equiv p_1-p_2$ (this $t$ is not to be confused with the one earlier in the entropic definition). It was noted in \cite{ref19} that there are only 9 independent parameters of $\chi$ because only three of the 9 elements of $T$ are independent, and can be written in terms of the vector $\vec{s}_{-}=\mathrm{tr}((p_1\rho_1-p_2\rho_2)*\vec{\sigma})$, with the relationship $T=\vec{e}\;\vec{s}^{\;T}_{-}$. Hence the three three-dimensional vectors $\vec{e},\vec{s_{\pm}}$ and $t$ constitute the 9 independent parameters that define a classical state. This formula for a classical state can be substituted into Eq. (\ref{eqn9}) and optimized with respect to the three vectors and parameter $t$, yielding the result for a geometric measure of quantum discord for two-qubits namely
\be\label{eqn10}
D^{(2)}_{A}(\rho)=\frac{1}{4}(\Vert x\Vert^2+\Vert T\Vert^2-k_{\mathrm{max}}).
\ee
Here $k_{\mathrm{max}}$ is the maximum eigenvalue of $xx^{T}+TT^{T}$. 

Consider classical states of $2 \otimes d$ systems. For $N$-qubit states, this corresponds to $d\equiv2^{N-1}$. These states are determined by $t\equiv p_1-p_2$, two angles defining the projectors $\Pi_i$ and two sets of $(d^{2}-1)$ parameters that define each density matrix. Note again that the local Bloch vectors are $3\times1$-, $(d^{2}-1)\times1$-dimensional and a correlation tensor that is $3\times(d^{2}-1)$-dimensional respectively. They are defined via the equations $x_{i}=\mathrm{tr}(\rho*\sigma_{i}\otimes I)$, $y_{i}=\mathrm{tr}(\rho*I\otimes O_{i})$ and $T_{ij}=\mathrm{tr}(\rho*\sigma_{i}\otimes O_{j})$, there being $(d^2-1)$ linearly independent operators $O_{i}$ defining an operator basis for the $d$-dimensional subsystem with the additional property $\mathrm{tr}(O_{i}O_{j})=\delta_{i,j}$. Here $\delta_{i,j}$ is the Kronecker symbol and a general density matrix is hence written in the Bloch representation as
\begin{widetext}
\be
\rho=\frac{1}{2d}\left(\openone+\displaystyle\sum_{i=1}^{3}x_{i}\sigma_{i}\otimes \openone+\displaystyle\sum_{j=1}^{d^2-1}y_{j}\openone\otimes O_{j}+\displaystyle\sum_{i=1}^{3}\sum_{j=1}^{d^2-1}T_{ij}\sigma_{i}\otimes O_{j}\right).
\ee
\end{widetext}
Again, it can be seen that the classical states can be defined in terms of three vectors and an additional parameter. This is done via $\vec{e}\equiv\mathrm{tr}(\langle\psi\vert\vec{\sigma}\vert\psi\rangle)$, and $\vec{s}_{\pm}\equiv\mathrm{tr}((p_1\rho_1\pm p_2\rho_2)*\vec{O})$ with $T=\vec{e}\;\vec{s}^{\;T}_{-}$. With $\Vert\vec{e}\Vert=1$, again another independent parameter is $t\equiv p_1-p_2$. The three parameters, $t$ and the three-dimensional unit vector $\vec{e}$, along with the two sets of $d^2-1$ parameters defining the two $d^2-1$-dimensional vectors $\vec{s}_{\pm}$, define all the elements of the three-dimensional vector $\vec{x}$, the $(d^2-1)$-dimensional vector $\vec{y}$ and the $3\times(d^2-1)$-dimensional tensor $T$. Thus, having established the Bloch representation of the classical state, once again, Eq. (\ref{eqn9}) can be optimized with respect to $t$, $\vec{e}$ and $\vec{s}_{\pm}$. Writing $\Vert\rho-\chi\Vert^2$ as
\begin{eqnarray}\label{distance}
\Vert\rho-\chi\Vert^2 &=& \Vert\rho\Vert^2+\Vert\chi\Vert^2-2\;\mathrm{tr}(\rho\chi).\\
&=&\frac{1}{2d}(1+\Vert\vec{x}\Vert^2+\Vert\vec{y}\Vert^2+\Vert T\Vert^2)+\nonumber\\
&+&\frac{1}{2d}(1+t^2+\Vert\vec{s}_{+}\Vert^2+\Vert\vec{s}_{-}\Vert^2)+\nonumber\\
&-&\frac{1}{d}(1+t\vec{x}\vec{e}+\vec{y}\vec{s}_{+}+\vec{e}T\vec{s}_{-}),
\end{eqnarray}
where $\Vert T\Vert^2\equiv\mathrm{tr}(T^{T}T)$. Once again, we have to optimize Eq. (\ref{distance}) over $\vec{s}_{\pm}$, $t$ and $\vec{e}$. We first optimize over the first two sets of parameters. As in the case of two qubits, since the Hessian is positive and non-singular, we compute the derivatives and find the minimum by setting them to zero. These equations yield
\begin{eqnarray}
\frac{\partial\Vert\rho-\chi\Vert^2}{\partial t}=\frac{1}{d}(-\vec{x}\vec{e}+t)=0,\\
\frac{\partial\Vert\rho-\chi\Vert^2}{\partial \vec{s}_{+}}=\frac{1}{d}(-\vec{y}+\vec{s}_{+})=0,\\
\frac{\partial\Vert\rho-\chi\Vert^2}{\partial \vec{s}_{-}}=\frac{1}{d}(-T^{T}\vec{e}+\vec{s}_{-})=0.
\end{eqnarray}
After substituting the solutions to these equations, which yield a global minimum, we get $\Vert\rho-\chi\Vert^2=(\Vert\vec{x}\Vert^2+\Vert T\Vert^2-\vec{e}(\vec{x}\vec{x}^{T}+TT^{T})\vec{e})/(2d)$. This is minimized by $\vec{e}$ being the maximum eigenvector of $(\vec{x}\vec{x}^{T}+TT^{T})$ corresponding to its maximum eigenvalue $k_{\mathrm{max}}$. The resulting minima again has the same form as Eq. (\ref{eqn10}) namely
\be
D^{(2)}_{A}(\rho)=\frac{1}{2d}(\Vert x\Vert^2+\Vert T\Vert^2-k_{\mathrm{max}}).
\ee
 This completes the proof that Eq. (\ref{eqn10}) represents he correct form for calculating geometric measure of quantum discord for  arbitrary $(2 \otimes d)$-dimensional systems. We note that this proof is completely analogous to the proof in \cite{ref19}. 

For $X$-states (whose elements have been defined according to \cite{ref12}), this formula can be written in a compact way. For two-qubit $X$-states, $\vec{x}=(0,0,d_{10})$ and 
\be
T=
\left(
\begin{array}{ccc}
a_{00}  & a_{01}  & 0  \\
 a_{10} & a_{11}  & 0  \\
0  &  0 &   d_{11}
\end{array}
\right).
\ee
which leads to the eigenvalues of $xx^T+TT^T$ to be $\lambda_{0}=d^2_{10}+d^2_{11}$  and 
\be
\lambda_{\pm}=\frac{1}{2}\left(\;\vert a\vert^2\pm\sqrt{\vert a\vert^4-\mathrm{a}^4}\;\right)
\ee
where $\vert a\vert^2=a^2_{00}+a^2_{01}+a^2_{10}+a^2_{11}$ and $\mathrm{a}^2=2(a_{00}a_{11}-a_{01}a_{10})$. So
\be
k_{\mathrm{max}}=\mathrm{max}(\lambda_{0},\lambda_{+})
\ee
This leads to the formula for $D^{(2)}_{A}(\rho)$ namely
\be
D^{(2)}_{A}(\rho)=\frac{1}{4}\{\vert a\vert^2+\lambda_{0}-\mathrm{max}(\lambda_{0},\lambda_{+})\}
\ee
which can be written compactly as
\be\label{gdx}
D^{(2)}_{A}(\rho)=\frac{1}{4}\mathrm{min}(\vert a\vert^2,\frac{1}{2}\vert a\vert^2+\lambda_0-\frac{1}{2}\sqrt{\vert a\vert^4-\mathrm{a}^4}).
\ee
which generalizes the formula given for states with maximally mixed marginals in \cite{ref19}. Note that this formula is the geometric analogue of the formulas discussed in \cite{ref7}, \cite{ref19} and \cite{ref31}. Note that Eq. (\ref{gdx})  involves the full seven parameters defining all two-qubit states defined up to local unitaries, whereas \cite{ref7} and \cite{ref19} involved three parameters and \cite{ref31} involved five parameters.

\section{Conclusions}\label{s4}
Constructing analytic formul{ae} for quantum discord is hard for two reasons. The first difficulty relates to being able to compute eigenvalues of matrices analytically, which is related to finding roots of polynomials. The second difficulty relates to the optimization over either the set of local measurements or over the set of classical states, depending on either the entropic or the geometric definition of discord. Here we present two formul{\ae}. The first formula involves a new class of states, namely extended $X$-states. All steps up to the optimization can be carried out analytically, the optimization itself then being at most over two parameters $(\theta, \phi)$. For density matrices of a certain off-diagonal structure, one parameter $\phi$ drops out, and even the remaining optimization in $\theta$ can be avoided for most density matrices to give an easily computed, useful formula as shown by numerical tests. The second formula established that an existing expression for the geometric measure of quantum discord for two qubits was valid for $(2 \otimes d)$-dimensional systems.

We note an alternative study of discord for qubit-qudit that used local operation and classical communication to reduce density matrices to a two-parameter set \cite{ref20}. Witnesses for discord in such states and a criterion called strong positive partial transpose for a non-zero discord have also been proposed \cite{ref21}. Bounds on a quantity called measurement-induced nonlocality that is dual to quantum discord were given for $2 \otimes d$ systems in \cite{ref22}. A related paper \cite{ref22p5} relies on \cite{ref22} to study arbitrary bi-partitions. The entanglement of formation for a certain class of $2 \otimes d$ systems were given in \cite{ref23} by starting with tripartite $2 \otimes 2 \otimes d$, and tracing over the last to end with an $X$ state for the $2 \otimes 2$ subsystem. Our study differs in considering directly a $2 \otimes d$ system, so long as it is of extended $X$ type, and evaluating discord  (according to two different expressions) without any further restriction on the parameters involved, numbering $8d-1$ for $d$ even or $8d-5$ for $d$ odd. Finally, after completion of our paper, we became aware of a paper \cite{ref29} that comes to many similar conclusions to ours for the two definitions of discord but, whereas their study is restricted to qubit-qubit, ours considers also qubit-qudit and $N$-qubits.      

\textit{Acknowledgements}: SV is supported by NSF under Project No. PHY -0902906.

\end{document}